**Properties of a one-dimensional periodicity of the gravitational interaction**

Francesco Scotognella
*Dipartimento di Fisica, Politecnico di Milano, Piazza Leonardo da Vinci 32, 20133, Milano, Italia*
*Email address: francesco.scotognella@polimi.it*

**Abstract**
*We briefly discuss the possibility to describe with a formalism, analogous to the Bragg law and the transfer matrix method used for photonic crystals, the behaviour of the kinetic energy of an object travelling through a one-dimensional (1D) modulation of the gravitational interaction, i.e. a 1D gravitational crystal. We speculate that certain ranges of the kinetic energy of an object with mass m and speed v cannot travel through the crystal, giving rise to a gravitational gap.*

In the paper by Griffiths and Steinke "Waves in locally periodic media" [1], the theory of wave propagation through 1D systems, has been reviewed in a very elegant way. Griffiths and Steinke have discussed different physical problems, as the optical properties of photonic crystals, and the electron wave functions in the quantum theory of solids [1]. To the best of our knowledge, systems in which a modulation of the gravitational interaction has never been studied.

In a 1D lattice as sketched in Figure 1, pairs of blocks of Aluminium (size 4x4x4 cm, mass $m_{Al}$=0.1729 kg) and Iron (size 4x4x4 cm, $m_{Fe}$=0.5039 kg) are alternated.

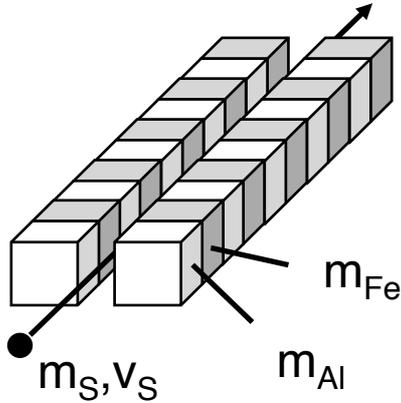

**Figure 1.** Scheme of 1D lattice where pairs of blocks of Aluminium and Iron are alternated. A sphere with mass $m_S$ and speed $v_S$ travels between the pairs of blocks (distance between the blocks of 1 cm, diameter of the sphere of 0.8 cm).

In this lattice a modulation of the gravitational interaction is obtained. In analogy with other 1D physical problems, we speculate that the kinetic energy, of a sphere of mass $m_S$=0.005 kg and speed $v_S$ interacting with the block of Aluminium and Iron, obeys a law that is similar to the Bragg law. Such law is of the form

$$E_K = 2(F_1 d_1 + F_2 d_2) \qquad (1)$$

with $F_1 = G m_S m_{Al}/d_{S-Al}^2$, $F_2 = G m_S m_{Fe}/d_{S-Fe}^2$, and $E_K = \frac{1}{2} m_S v_S^2$. G is the gravitational constant (6.67×10⁻¹¹ Nm²/kg²).

In a transfer matrix method, analogous to the one used for photonic crystals [2],

$$M_j = \begin{bmatrix} cos\phi_j & -\frac{i}{m_j} sin\phi_j \\ -i m_j sin\phi_j & cos\phi_j \end{bmatrix}, \qquad (2)$$

with $\phi_j(E) = E/(\xi F_j d_j)$, is the characteristic matrix of each layer. $\xi$ is a parameter (its value is 3.5). From the product of all the matrixes corresponding the layers of the one-dimensional lattice

$$M = \prod_j M_j \qquad (3)$$

we can determine the transmission coefficient for the 1D lattice, that is

$$t = \frac{2}{M_{11} + M_{12} + M_{21} + M_{22}} \qquad (4)$$

and gives rise to the transmission $T = |t|^2$.

In Figure 2 the calculated transmission spectrum, for the lattice depicted in Figure 1, is shown. The first and the second order of the gap are observed.

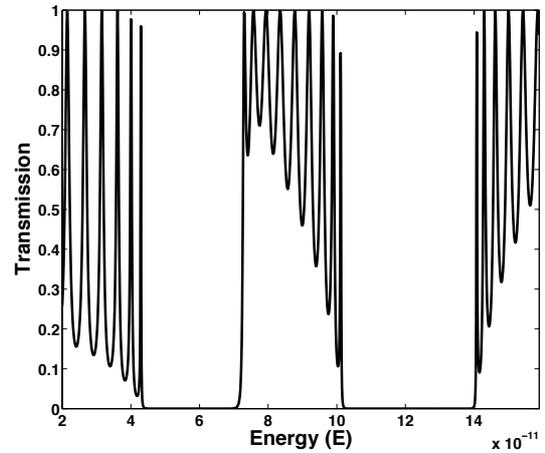

**Figure 2.** Transmission spectrum of a 1D lattice made by the alternation of Aluminium and Iron.

The centre of the gap for the lattice is at about 5.78×10⁻¹¹ J, corresponding to a speed of the sphere of about 1.52×10⁻⁴ m/s (a result is in agreement with Equation 1).
A possible development of this work is the accurate derivation of the formalism.
The experimental observation of the suggested phenomenon can be achieved in three ways: i) the study of systems with very small values of speed; ii) the study of systems with high volumetric mass density; iii) the highly precise fabrication of the lattice in order to observe the higher orders of the gap (corresponding to higher values of speed).